# Three-dimensional microfabrication through a multimode optical fiber

**Authors**:

**Edgar E. Morales-Delgado**
Laboratory of Applied Photonics Devices, School of Engineering, École Polytechnique Fédérale de Lausanne (EPFL), Station 17, 1015, Lausanne, Switzerland
*edgar.moralesdelgado@epfl.ch*

**Loic Urio**
Laboratory of Applied Photonics Devices, School of Engineering, École Polytechnique Fédérale de Lausanne (EPFL), Station 17, 1015, Lausanne, Switzerland
*loic.urio@epfl.ch*

**Donald B. Conkey**
Laboratory of Optics, School of Engineering, École Polytechnique Fédérale de Lausanne (EPFL), Station 17, 1015, Lausanne
*donald.conkey@epfl.ch*

**Nicolino Stasio**
Laboratory of Optics, School of Engineering, École Polytechnique Fédérale de Lausanne (EPFL), Station 17, 1015, Lausanne
*nicolino.stasio@epfl.ch*

**Demetri Psaltis**
Laboratory of Optics, School of Engineering, École Polytechnique Fédérale de Lausanne (EPFL), Station 17, 1015, Lausanne, Switzerland
*demetri.psaltis@epfl.ch*

**Christophe Moser\***
Laboratory of Applied Photonics Devices, School of Engineering, École Polytechnique Fédérale de Lausanne (EPFL), Station 17, 1015, Lausanne, Switzerland
*christophe.moser@epfl.ch*
Telephone number : +41216936110

\*corresponding author

## Abstract

Additive manufacturing, also known as 3D printing, is an advanced manufacturing technique that allows the fabrication of arbitrary macroscopic and microscopic objects. All 3D printing systems require large optical elements or nozzles in proximity to the built structure. This prevents their use in applications in which there is no direct access to the area where the objects have to be printed. Here, we demonstrate three-dimensional microfabrication based on two-photon polymerization (TPP) with sub diffraction-limited resolution through an ultra-thin, 50 mm long printing nozzle of 560 µm in diameter. Using wavefront shaping, femtosecond infrared pulses are focused and scanned through a multimode optical fiber (MMF) inside a photoresist that polymerizes via two-photon absorption. We show the construction of arbitrary 3D structures of 500 nm resolution on the other side of the fiber. To our knowledge, this is the first demonstration of microfabrication through a multimode optical fiber. Our work represents a new area which we refer to as endofabrication.

## INTRODUCTION

3D printing technologies comprise several fabrication methods that allow the direct construction of objects of arbitrary shape and size made of materials such as plastics, metals, and ceramics. Direct Laser Writing (DLW) is one of the 3D printing methods that enable the fabrication of high resolution microstructures. The resulting structures are made of a broad selection of polymer-based materials of various mechanical, electrical and optical properties. The applications range from scaffold-like structures for cell growth [1, 2] to optical components such as complex micro lenses, waveguides, diffractive optical elements, photonic crystals, optical data storage, microfluidics, nanophotonics or micro-electromechanical systems (MEMS) [3-15]. All these structures can be fabricated with hundreds of nanometers in resolution.

The principle behind DLW is based on photo-polymerization, a phenomenon in which a photosensitive material reduces its solubility and becomes solid when exposed to light. Such materials are based on a resin consisting of chemical compounds known as monomers which, in combination with a photo-initiator, become cross-linked upon absorption of photons producing a subsequent polymerization inside the illuminated volume [16]. As in fluorescent microscopy, the photo-polymerization phenomenon can also be triggered by multi-photon absorption processes. In the case of two-photon polymerization (TPP), its nonlinear dependence on peak intensity results in a tightly confined volume of both absorption and polymerization within the high energy focused pulses impinging in the photoresists, resulting in a smaller polymerized volume when compared to single photon polymerization. Scanning of the light focus with respect to the photoresists allows the fabrication of structures [10, 17-20]. A post exposure phase known as "development" dissolves and removes the unexposed portion of the photoresists.

First achieved by Mauro et al. in 1996 [18], TPP has since then reached resolutions smaller than 100 nm [21]. As in STED microscopy, where a second light beam can be used to prevent fluorescence around the excitation light focus, a second beam can be used to inhibit polymerization in the surroundings of the light focus used to polymerize, allowing the construction of micro structures with hundreds of nanometers in resolution in a single photon absorption process [22, 23].

The main drawback of all the DLW methods is that the 3D printed structures can only be created in proximity to very large optical elements. This prevents the microfabrication of objects inside the body or in cavities difficult to access. Here we propose a new method that overcomes those limitations.

In parallel to the progress of photo-polymerization, several techniques have been developed to focus and scan light patterns [24-39] and optical pulses [40-44] through scattering media and multimode optical fibers. Different than a lens, the multimode fiber itself does not directly allow the transmission of images or focused optical pulses. In the fiber, dispersion and modal mixing, given by the restricted propagation of light as a combination of a large but limited number of modes, produces a scrambled intensity profile and a broad temporal profile. We have previously developed a method that allows the transmission of focused optical pulses with a femtosecond temporal profile unaffected by modal dispersion [42].

In this paper, we demonstrate experimentally two-photon polymerization 3D printing through a fiber-based optical system. First, using a very simple setup, we characterize the minimum power and exposure time that allows polymerization of the photoresists with a microscope objective to investigate the feasibility of photo-polymerization under similar light focusing conditions as those available when focusing light through a multimode optical fiber. Then, using a method for scanning and focusing of optical pulses through multimode fibers based on optical phase conjugation [45, 46] and wavefront shaping [42], fiber modes with the same group velocity are excited and forced to interfere at a given location forming a high intensity optical pulse. This focused pulse is scanned in a 3D volume of a photoresist, producing polymerization within the optical voxel, allowing the fabrication of structures of arbitrary shape with sub diffraction-limited resolution through the multimode fiber. This scanning method does not require any distal scanning element, allowing an ultra-thin 3D printing probe.

From a technological perspective, the proposed ultra-thin 3D printer enables high resolution microfabrication in a whole new set of applications where conventional 3D printing methods cannot have access, such as inside of biological tissue or in the interior of already built structures and places with difficult access.

## MATERIALS AND METHODS

Our experiments were conducted using a Ti:sapphire laser (Coherent Mira 900; with central wavelength λc=800 nm; spectral width σλ=10 nm; pulse length of 100 fs; repetition rate 76 MHz and maximum output average power of 800 mW). The photoresist used in all our experiments is IPL 780 from Nanoscribe GmbH, which can be polymerized via two-photon absorption at an excitation wavelength centered at 800 nm.

### Two-photon polymerization through a microscope objective

In order to establish the feasibility of multimode fiber 3D printing, we begin with simple DLW experiments, which consist on producing two-photon polymerization under peak power conditions as close as possible to those expected when focusing light through multimode fibers. To meet such conditions, we use a 40x microscope objective of 0.65 NA to focus the femtosecond light beam on the photoresists as shown in Fig. 1 (a). For DLW, objectives with very high NA are typically used (oil immersion objectives with NA = 1.4). However, in this case we use a lower NA objective because our goal is to polymerize through a lensed multimode fiber with a numerical aperture equal to 0.54. A detailed explanation of the DLW experiments is presented in the section S2 of the Supplementary Information.

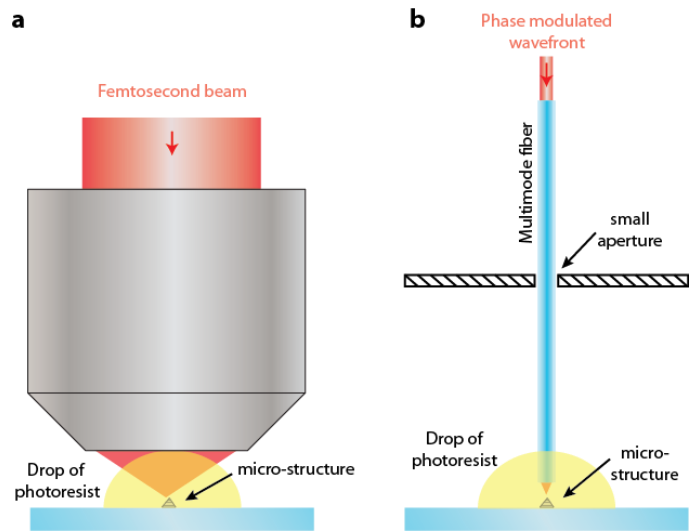

Figure 1. Working principle of direct laser writing (DLW). (a) DLW based on a microscope objective. (b) DLW based on a multimode fiber. The fiber can enter and micro-fabricate inside areas difficult to access.

## Focusing and scanning of femtosecond pulses through the multimode fiber

As explained in the introduction, a beam focused on one side of a multimode optical fiber produces a speckle like pattern on the other side as shown in Fig. 2 (a). Therefore, to focus and scan optical pulses through the multimode fiber, we use time-gated digital phase conjugation [41, 42], which consists of two steps: a calibration step that obtains the phase that can be modulated to generate focused spots on the other side of the fiber followed by a reconstruction step that reproduces the phases and generates the spots through the fiber. These two steps are explained in detail in section S3 of the Supplementary Information. The fiber is graded index and has a length of 50 mm and a core diameter of 400 µm.

The focused pulses generated through the MMF, shown in Fig 2 (c), are 800 times more intense than the average background, with a two-photon spot contrast of 640'000, reducing any possibility that the background intensity will produce any polymerization.

The average pulse length of the pulses focused through the fiber is 115 fs. A sample of those traces is shown in Fig. 2 (d). This value is shorter than the 210 fs pulse length that results when no wavefront control is used. This broader pulse is shown in Fig. 2 (b). The pulse width measurements are explained in the section S4 of the Supplementary Information.

The maximum average power that our system can deliver through the fiber is 12 mW within the central part of the Airy disk, with a peak power of 55 $GW/cm^2$ corresponding to 6 $mJ/cm^2$ and a pulse energy of 160 pJ.

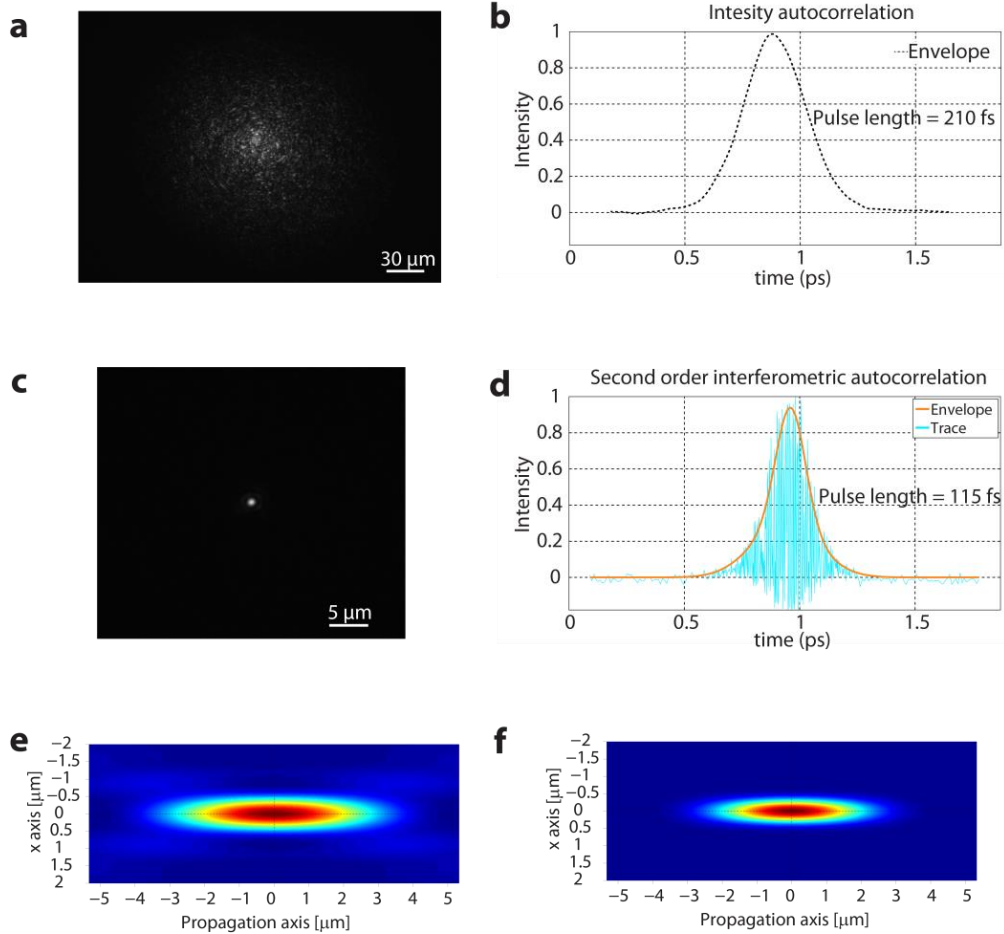

Figure 2. Pulsed light transmission through the multimode fiber. (a) Speckle-like pattern produced on one end of the fiber when light is focused with a microscope objective on the other end. (b) Intensity autocorrelation of the speckle like pulse shown in (a). Pulse length is 210 fs. (c) Intensity pattern when light is focused through the fiber using spatial light modulation. (d) Second order interferometric autocorrelation of the spot shown in (c). Pulse length is 115 fs. (e) One-photon 2D PSF of the pulse delivered through the fiber (simulation). (f) Two-photon 2D PSF of the pulse delivered through the fiber (simulation).

## Two-photon polymerization through the multimode fiber

Once the calibration is completed, the system is able to polymerize over the field of view (FOV) at a 40 µm working distance from the fiber tip. We select this working distance to prevent the polymerized structure from being attached to the tip of the fiber. The spot is scanned in this plane digitally by changing the phases on the SLM. The vertical scan is performed by a motorized stage that lowers the sample after polymerizing each plane. We use the dip-in arrangement show in Fig 1 (b) in which the fiber is dipped into the photoresist. The photoresist is previously bubbled with nitrogen to remove the oxygen that prevents polymerization, as explained in the section S1 of the Supplementary Information.

To characterize the polymerization capabilities of our device, we polymerize a grid of voxels separated by 6 μm over a 108 μm diameter FOV. The sample is scanned vertically with a 2 um step, "printing" one layer of voxels on top of the previous one. The printed rods are shown from above in Fig 3 (a, b), taken with a differential interference contrast microscope (DIC microscope). We print the voxels starting 4 um inside the glass to ensure the adhesion of the printed structures to the glass substrate even if the sample is slightly tilted.

We measure the full width at half maximum (FWHM) size of the printed rods. When focusing light through multimode fibers the numerical aperture depends on the focal position [30]. Additionally, the presence of the graded index lens introduces third order aberrations which become more significant away from the optical axis [42]. This produces a radial dependence of generated spot intensity as a function of position, which in turn leads to a radial dependence of the voxel diameter as shown in Fig. 3 (a). Since the exposure dose also affects the voxel size, we perform a radial compensation of exposure in order to equalize the voxel size over the whole field of print (FOP) as shown in Fig 3 (b, c). The small slope of the continuous curve in Fig. 3 is produced by the inherent limitation of our phase-only SLM being unable to imprint phase masks faster than a 20 Hz refresh rate, resulting in a non-optimal voxel size compensation for small radii.

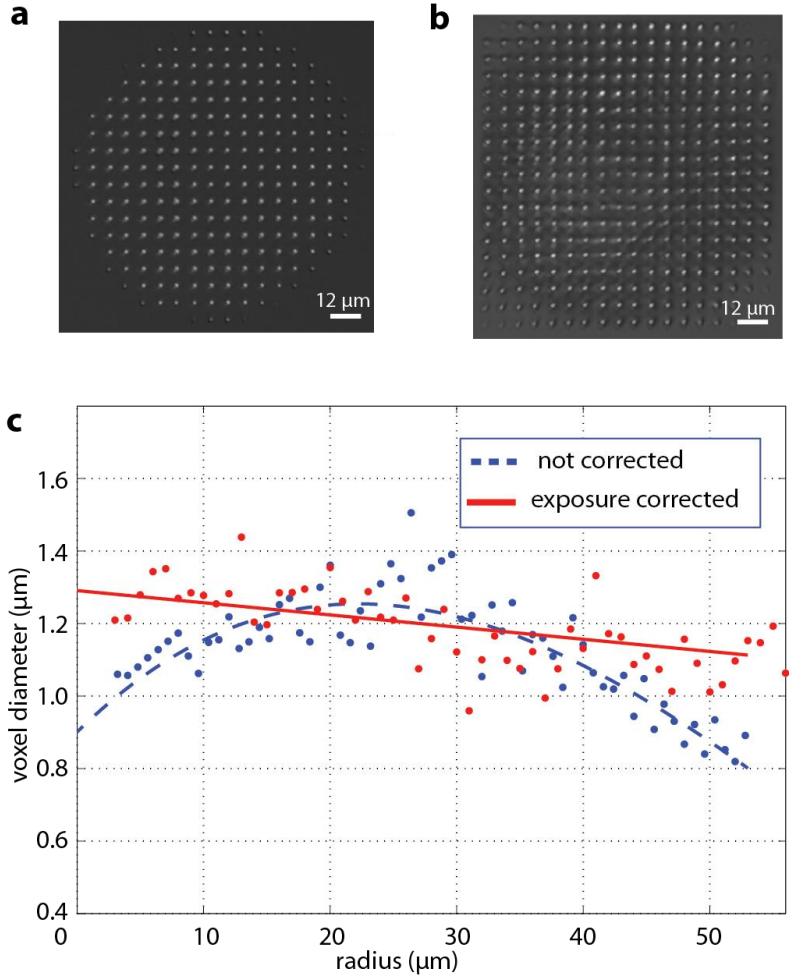

Figure 3. Characterization of the 3D printing system. (a) Grid of voxels printed through the multimode fiber. (b) Grid of voxels printed through the multimode fiber in the case of exposure time correction. A radial correction of exposure is applied to polymerize voxels of uniform size. (c) Voxel diameter versus radial position for (a) and (b).

To measure the voxel diameter dependence on exposure, we build the same structure at different exposure times as shown in Fig. 4 (a). The voxel diameter versus exposure time curve shown in Fig. 4 (c) follows a logarithmic form which corresponds to the exponential decay of the concentration of monomers upon light exposure [47]. At longer exposures, monomers are consumed around the voxel area preventing its growth. The minimum exposure time that we can apply is limited by the refresh rate of the SLM, which is quite low (20 Hz).

To find the smallest spot size without having the slow SLM refresh rate restriction, we designed the following experiment: A circle of 12 voxels at the same radial position of 15 µm separated by an angle of 30° is printed through the fiber. As explained above, without any exposure time correction, the voxel size depends on the radial position. Therefore, in order to keep the same intensity for all the voxels, we build the voxels in a circular arc at a fixed radius. The exposure time per diffraction-limited spot is controlled by setting the motorized stage to different speeds. This way, we can identify the threshold exposure time required for polymerization. The results are shown in Fig 4 (c). The smallest voxel size that we were able to polymerize is 400 nm FWHM, which corresponds to a printing resolution of 480 nm. This is smaller than the optical resolution due to the non-linearity of the cross-link process occurring in photopolymerization. We clarify that all these measurements where performed before developing the sample, which means that we are not "taking advantage" of the shrinkage effect produced when developing the sample, that can lead to an even better voxel resolution. The shrinkage effect however, is complex to control precisely [48].

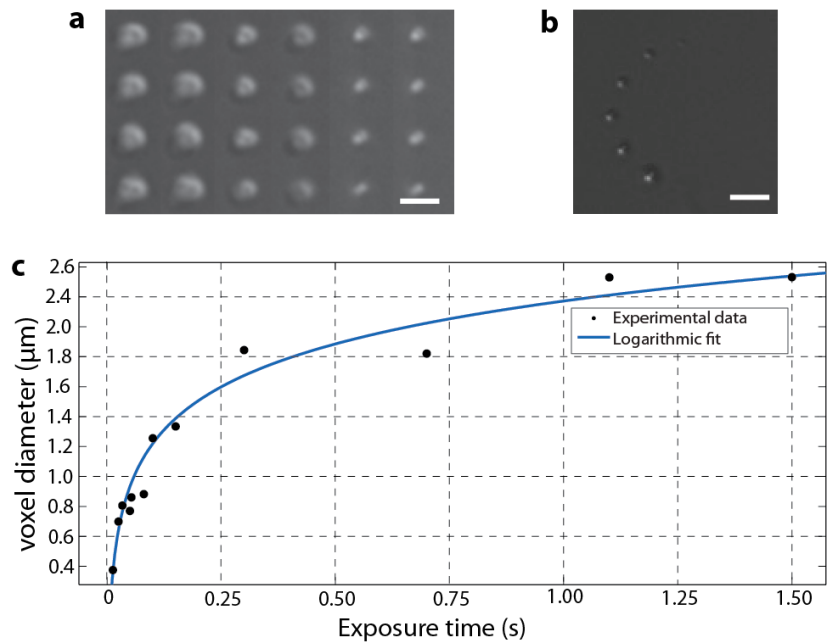

Figure 4. Voxel dependence on exposure time. (a-b) Example of voxels printed with 1.5, 1.1, 0.7, 0.3, 0.15, 0.1, 0.05, 0.03, 0.025, 0.0125 seconds of exposure. Scale bars are 7.5 µm. (c) Voxel diameter dependence on exposure time. The logarithmic behavior arises due to an exponential decay of monomer concentration when the photoresists is exposed to light.

## RESULTS AND DISCUSSION

### The multimode fiber 3D printer

Once we are able to polymerize with a uniform voxel resolution of 500 nm over the field of print, we can create arbitrary 3D structures. For instance, we created a model of the Pyramid of Chichén Itzá of 100 µm x 100 µm base size shown in Fig. 5. Figure 5 (a) shows a DIC image of the undeveloped model. Figures 5 (b, c) show the scanning electron microscope (SEM) images of the developed model. The lateral printing step used is 1 µm and the axial 2 µm, resulting in a lattice type surface. The time required to print this structure was 25 minutes approximately. The printing speed is limited by the refresh rate of the used phase-only SLM (20 Hz).

To obtain the SEM images, the sample is developed to remove the unexposed photoresist. This is achieved by submerging the sample in Propylene glycol methyl ether acetate (PGMEA) during 10 minutes, which is a common developer used in photolithography. Then, the sample is washed with isopropanol. Finally, the sample is coated with a 4 nm gold layer in preparation for the SEM.

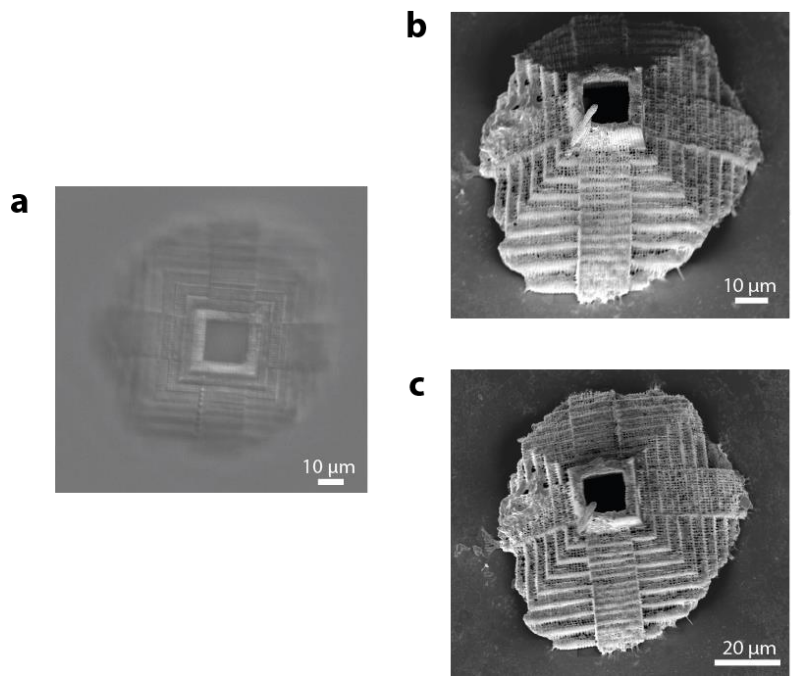

Fig. 5. Model of the Pyramid of Chichén Itzá 3D printed through a multimode optical fiber. The base diameter is as small as the thickness of a human hair. (a) Image of the pyramid acquired with a differential interference microscope. (b,c) Images of the pyramid acquired with a scanning electron microscope.

There are several methods in addition to time gated phase conjugation to accomplish focusing through turbid media or multimode fibers which can be employed, such as, the measurement of the transmission matrix or iterative methods. In this particular case of multiphoton polymerization through multimode fibers, the power output available in our femtosecond laser gave a peak power at the sample very close to the threshold required for two-photon polymerization. In fact, the threshold average power of 10 mW experimentally measured in our system is close to the 12 mW of maximum average power in the multimode fiber printing system. This means that for the projection of more complex light patters or simultaneous spots through the fiber, generated using the transmission matrix method, a larger power would be required to actually polymerize. Therefore, we used the method of time-gated digital phase conjugation, which is simple but sufficient to generate sequential light foci through the fiber.

The whole potential of the transmission matrix method in 3D printing can be fully exploited in cases where a larger power budget is available, such as in single-photon polymerization.

## CONCLUSIONS

We have presented the first demonstration of additive manufacturing through an ultra-thin probe of 560 µm diameter, 50 mm length with a voxel resolution of 500 nm over a 110 µm diameter field of print. Our method does not have any intrinsic restriction to the height of the printed structure.

The presented 3D printing probe based on a multimode optical fiber does not contain any distal scanning element. Printing is achieved via two-photon polymerization. A focused ultrashort pulse scanned digitally with a SLM from the proximal side can print any arbitrary structure on the other end of the fiber.

The maximum power that we are able to deliver through the fiber is close to the threshold power for polymerization. Therefore, we could only print one voxel at a time. In principle, with a more powerful laser source, we could reconstruct on the SLM simultaneous plane waves either by addition of waves at the SLM point or by using the TM, enabling a faster printing speed, reducing the printing time by a factor of N, where N is the number of simultaneous spots generated through the fiber.

This pioneering work enables a new area that we call endofabrication, which consists on the fabrication of micro-structures inside places that are difficult to access by conventional 3D printing devices and that are accessible only through very small apertures.

## ACKNOWLEDGEMENTS

We would like to thank Dr. Miguel A. Modestino from the Laboratory of Optics and the Laboratory of Applied Photonics Devices at EPFL for recording the SEM images. We also thank Dr. Michael Thiel for providing pertinent information about the Nanoscribe IPL-780 photoresist and the sample development procedure. We acknowledge the funding research grant SNF with reference number: 200021_160113 (MuxWave).

# SUPPLEMENTARY INFORMATION

## S1. Principle of two-photon polymerization

Photoresists for TPP are commonly composed of monomers mixed with photoinitiators. The monomers are molecules that can form polymer chains by cross-linking [16], making the cross-linked regions of the photoresist less soluble to a developer. Polymerization and cross-linking is triggered by the presence of free radicals [49, 50] whose generation can be light-induced. The molecules that generate free radicals upon light exposure are the photoinitiators.

An element present in the air that can inhibit polymerization is oxygen, which reacts with the radicals generated by the exposed photoinitiator to form peroxy radicals that does not trigger cross-linking [51]. At high light intensities, the generation rate of radicals from the exposed photoinitiator is larger than the oxidation rate and polymerization will occur. Otherwise, for low light intensities, oxygen diffusion from the surrounding air prevents any polymerization [16] as the generation rate of radicals is smaller than the oxidation rate produced by oxygen diffusion.

In polymerization the volume where the photoresist is polymerized is called a voxel. Being able to achieve the smallest possible voxel size is crucial to create precise micro and nanoscale 3D structures. As mentioned before, oxygen diffusion, exposure dose and optical resolution are the quantities that set the minimum achievable voxel size and the power threshold for polymerization.

Another important characteristic of photopolymerization is the inherent threshold behavior of the cross-linking process. If the intensity of the excitation light is above a certain threshold value, the exposed volume is polymerized. Otherwise, polymerization does not occur. This threshold characteristic happens due to the non-linear solubility of the photo-resist in the developer as a function of exposure dose [16]. The solubility drops at a given exposure dose, leading to the threshold behavior depicted in Fig. 2 (a). In this figure, the intensity distribution corresponds to the light intensity at the focal plane of a microscope objective (line profile of an Airy disk).

The exposure dose can be expressed as:

$$D \propto \tau P^N \quad (1)$$

Where $\tau$ is the exposure time, $P$ the power and $N$ the non-linearity (or linearity) of the absorption mechanism. As shown by Mueller et al. in [16], the threshold power $P_{thres}$ is proportional to $\tau^{-1/N}$, as solved from Eq. 1, but only for small exposure times, for which the threshold power is bigger than in the case of long exposures. For longer exposure times, a similar exposure dose can be obtained by having a lower power. However, the lowering of the power needed for polymerization has a limit set by the radicals that produce quenching or inhibition of the polymerization. For low light intensities, the generation rate of radicals is smaller than the oxygen diffusion rate and polymerization is not possible. Therefore, at low excitation powers, the power threshold is set by oxygen diffusion rather than by the degree of cross-linking necessary to polymerize.

All the previous considerations about photopolymerization hold true whatever the excitation process is (one or two-photon excitation). The photopolymerization process occurs as long as the light intensity possesses enough energy to excite the photoinitiator to produce radicals that then react with the monomer of the photoresist to cross-link it.

## S2 Two-photon polymerization through a microscope objective

Fo the DLW experiment with a microscope objective, the beam waist is set to cover the back aperture of the objective to achieve a diffraction limited focus (first Airy disk radius of 750 nm as shown in Supplementary Fig. 1 (a)).

As explained in section S1 of the Supplementary Information, the voxel diameter depends on the time the photoresists is exposed to light. Therefore, in our first experiment, we characterize the linewidth dependence on exposure time. The photoresist is sandwiched and sealed between two coverslips as illustrated in Supplementary Fig. 1 (b). The excitation is delivered through a 170 µm thickness glass which reaches the photoresists at the flat interface. We set an average power at the back aperture of the objective at 20 mW, which is close to the peak power available when polymerizing through the multimode fiber. To achieve different exposures, we move the microscope objective at different velocities with respect to the photoresists along an 80 µm line. The focused light on the photoresists follows the path shown in Supplementary Fig. 1 (c) to ensure illumination of the photoresists even if the sample is slightly tilted with respect to the movement of the objective. To perform a linewidth characterization without shrinkage [47], we measure the linewidth with a differential interference contrast microscope (DIC microscope) without developing the sample. A typical DIC image of those samples is shown in Supplementary Fig. 1 (d). The results are shown in Supplementary Fig. 1 (e). The maximum writing speed for which polymerization was detected is 90 µm/s, which corresponds to an exposure time per diffraction-limited spatial focus of 8.3 ms. The linewidth decreases for higher writing speeds as expected, with the smallest measured linewidth equal to 1.9 µm. The linewidth is thick due to the layered structure used as illustrated in Supplementary Fig 1 (c), which over exposes several times the polymerized voxels when writing the next layer.

As mentioned in the section S1 of the Supplementary Information, oxygen present in the air diffuses into the photoresist and acts as a cross-link inhibitor. In order to decrease the concentration of oxygen, nitrogen can be bubbled into the recipient that contains the photoresist, lowering the threshold power required for photopolymerization [16]. Hence, for the next experiment, nitrogen is bubbled into the photoresist during 10 minutes. The sample is exposed with the 40x objective at different velocities and at decreasing powers until identifying the minimum power that produces polymerization. Supplementary Fig. 1 (f) shows the resulting threshold powers at different writing speeds for both cases: with and without nitrogen bubbling. For low writing speeds, a plateau gives a lower threshold power limit, for which no matter how long the exposure is, no polymerization will occur. For writing speeds above 100 µm/s, the threshold power increases with writing speed.

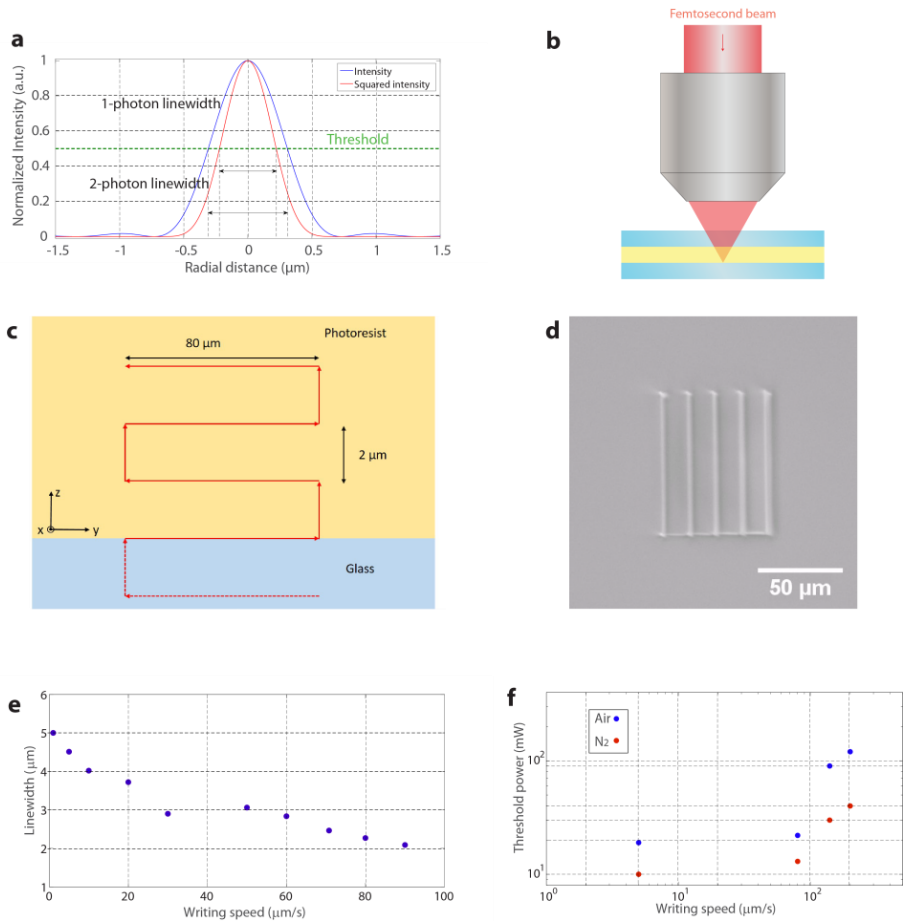

Supplementary Fig. 1. Two-photon polymerization with a 40x microscope objective (NA 0.65). (a) Calculated PSFs (one and two-photon ). (b) DLW arrangement. (c) Lateral view of the printed lines. (d) Top view of the printed lines. (e) Linewidth dependence on writing speed. (f) Threshold power dependence on writing speed with and without the presence of oxygen.

After nitrogen is bubbled, the threshold average power can be decreased by half (10 mW) compared to the case in which no nitrogen is used (20 mW) as shown in Supplementary Fig. 1 (f) for all writing speeds. Since our method for focusing femtosecond pulses through multimode fibers relies on a diffractive element (a spatial light modulator) and suffers from fiber coupling losses, a lower threshold power for photopolymerization is desired. Hence, for all subsequent fiber experiments, nitrogen was bubbled into the photoresist prior to polymerization.

## S3. Focusing and scanning of femtosecond pulses through the multimode fiber

In the calibration step the "calibration beam" (see Supplementary Fig. 2 (a)) is focused by a 60x (OBJ2) objective 40 μm in front of one end of the MMF. We calibrate with the photoresist between the fiber tip and the objective to take into account the aberrations produced by the photoresist. Then, light is coupled into the fiber generating a time varying speckle patter on the proximal side, which is imaged by a 20x objective (OBJ1) and a 150 mm lens (L1) on a CCD. On the CCD an off-axis hologram of the field is recorded with the pulsed reference beam, which can be delayed by a motorized stage to the delay position that results in the largest visibility of the fringes [41]. Then a digital hologram is recorded. This hologram contains the phase and amplitude information of the time sampled field emerging from the fiber at the time set by the gating reference.

In the reconstruction step (Supplementary Fig. 2 (b)), the flip mirror is disabled allowing the reference to reconstruct on a spatial light modulator (SLM, Holoeye Pluto, 1920x1080) the calculated phase conjugated replica of the field recorded in the digital hologram. From the SLM, light retraces back through the optical setup and the multimode fiber interfering constructively on the distal side forming a bright intensity focus at the location of the calibration spot. We repeat the calibration and reconstruction process several times with OBJ2 at different locations in order to store the holograms that allow the reconstruction of spots in the plane 40 um away from the tip of the fiber.

In a multimode optical fiber the polarization state of a circularly polarized beam is preserved upon propagation, which is not held in the case of linearly polarized fiber excitation [32]. In practice, this is implemented by the quarter wave plates (λ/4) show in Supplementary Fig. 2. The first one transforms the linearly polarized beam into a circularly polarized beam and at the other fiber end it transforms the circularly polarized beam coming from the fiber into a linearly polarized beam with same polarization orientation as that of the linearly polarized reference beam used to record the hologram. With this implementation in place, we measured an enhancement of 25% in the spot intensity to background ratio.

When focusing light through non-imaging optical elements such as scattering media or multimode optical fibers, the capability to form a sharp bright intensity focus depends on the number of degrees of freedom that can be controlled [25]. In the case of the multimode fiber, it depends on the number of modes that can coherently interfere simultaneously through the fiber, which in this case depends on the fiber's diameter, fiber NA and fiber length. Graded index fibers exhibit a smaller temporal multimode broadening when compared to step index fibers [52]. The fiber chosen for our experiments is a commercially available graded index fiber from Fiberware GmbH with a 400 µm core diameter, 560 µm cladding diameter, a length of 50 mm and a NA equal to 0.29.

In order to enhance the resolution of the 3D printing system we attached a graded index lens of 350 µm diameter to one end of the multimode fiber as shown in the fiber of Supplementary Fig. 2 (a). This increases the numerical aperture of the lensed side of the fiber at the optical axis from 0.29 to 0.54, and reduces the field of view (FOV) of the fiber from 400 µm to 210 µm. An alternative is to build a lens directly on the tip of the fiber as in [53, 54].

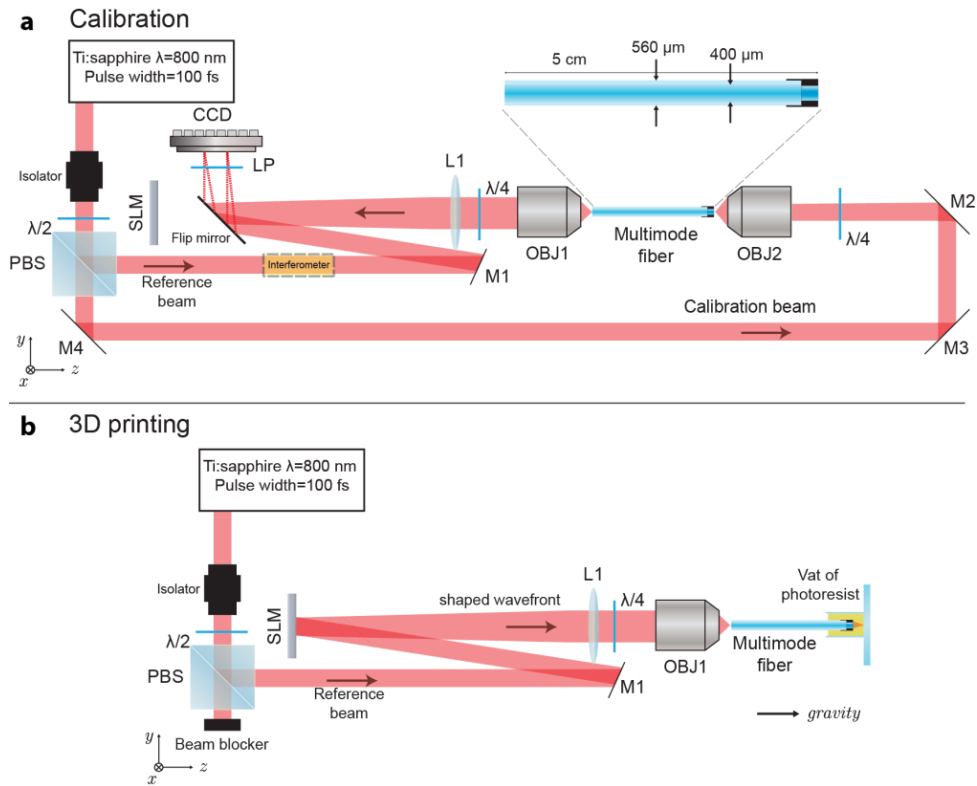

Supplementary Figure 2. Experimental setup. (a) Calibration. (b) Reconstruction for 3D printing.

### S4. Pulse width measurement

The pulse lengths of the focused pulses generated through the fiber are measured using second order interferometric autocorrelation. This operation requires the introduction of two replicas of the pulses on the reference arm, which is performed using an interferometer show in Supplementary Fig. 2 (a). The nonlinearity detection required for this measurement is introduced by focusing the two replicas of the focused pulses in a two-photon fluorescent sample of PDMS mixed with Rhodamine 6G, which gives the same result as if a second harmonic generation crystal is used [55, 56]. The fluorescent signal propagates through the fiber and is detected by a PMT for each one of the different discrete temporal overlaps between the two replicas of the pulses, resulting in an autocorrelation trace like the one shown in Fig. 2 (d). To calculate the value of the pulse length, the $sech^2$ pulse shape of the optical pulses is taken into account. For the pulse width measurement in which there is not wavefront control shown in Fig. 2 (b) the pulse length is measured using an intensity autocorrelator (Carpe, APE).